# Three-point functions in noncompact lattice QED [*]

M. Göckeler[a,b], R. Horsley[a], P.E.L. Rakow[c] and G. Schierholz[a,d]

[a]Höchstleistungsrechenzentrum HLRZ, c/o KFA, Postfach 1913, D-52425 Jülich, Germany

[b]Institut für Theoretische Physik, RWTH Aachen, Physikzentrum, D-52056 Aachen, Germany

[c]Institut für Theoretische Physik, Freie Universität Berlin, Arnimallee 14, D-14195 Berlin, Germany

[d]Deutsches Elektronen-Synchrotron DESY, Notkestraße 85, D-22603 Hamburg, Germany

We calculate fermion-antifermion-"meson" three-point functions in noncompact lattice QED with dynamical staggered fermions and use them to extract effective Yukawa couplings. The results are consistent with the hypothesis that QED is trivial.

In strongly coupled lattice QED chiral symmetry is spontaneously broken, whereas for small bare charges one expects an unbroken chiral symmetry [1]. This observation naturally leads to the question if a nontrivial continuum limit can be defined at the corresponding critical point. Assuming that this is the case one would also like to know the nature of the associated continuum theory. To investigate these problems several groups have performed extensive Monte Carlo simulations of lattice QED using various techniques (see e.g. [2–7]). Yet a consensus has not been reached (see also [8]).

From correlation functions of composite fermion-antifermion operators one finds at least two fermion-antifermion (bound) states ("meson" states): a pseudoscalar state (P, the Goldstone boson associated with the spontaneous symmetry breaking) and a scalar state (S) [5,9,10]. (Effective) Yukawa couplings of these states to the fermions can be extracted from suitable three-point functions, whose investigation is the subject of this talk. The study should provide additional information about the model. In particular, if the theory is trivial, the couplings should vanish as one approaches the critical point.

We have performed simulations with dynamical staggered fermions using the noncompact formulation of the gauge field action. The total action is then given by $S = S_G + S_F$ with

$$S_G = \tfrac{1}{2}\beta \sum_{x,\mu<\nu} F_{\mu\nu}^2(x), \qquad (1)$$

$$S_F = \sum_x \left\{ \tfrac{1}{2} \sum_\mu \eta_\mu(x) \left( \overline{\chi}(x) \mathrm{e}^{\mathrm{i}A_\mu(x)} \chi(x+\hat{\mu}) - \overline{\chi}(x+\hat{\mu}) \mathrm{e}^{-\mathrm{i}A_\mu(x)} \chi(x) \right) + m \overline{\chi}(x)\chi(x) \right\} \qquad (2)$$

where $\beta = 1/e^2$ ($e$ = bare charge) and

$$\eta_\mu(x) = (-1)^{x_1+\cdots+x_{\mu-1}}, \qquad (3)$$

$$F_{\mu\nu}(x) = A_\mu(x) + A_\nu(x+\hat{\mu}) - A_\mu(x+\hat{\nu}) - A_\nu(x). \quad (4)$$

We work on a $12^4$ lattice choosing periodic boundary conditions for the gauge field $A_\mu$ and periodic (antiperiodic) spatial (temporal) boundary conditions for the fermion field $\chi, \overline{\chi}$.

Fermionic observables like the fermion propagator or the three-point functions to be studied need gauge fixing. We work with the Landau gauge, which can be implemented exactly in the noncompact formulation. Since the zero-momentum mode of the gauge field does not average to zero in our ensembles, we form blocks of, e.g., 20 successive configurations, on which this mode is approximately constant, and perform our analysis in each block separately taking the zero-momentum mode into account explicitly (cf. [10]). From these block results we then calculate the averages and errors.

---

[*]Talk given by M. Göckeler at LATTICE 93

According to the triviality scenario, the renormalized charge vanishes in the continuum limit. Furthermore, the chiral phase transition is well described by logarithmically improved mean field theory, i.e., the equation of state can be derived from an effective linear $\sigma$-model [10]. In particular, the behaviour of the effective quartic coupling $\lambda$ is governed by a (one-loop) renormalization group equation of the form

$$\frac{d\lambda}{dt} = c\lambda^2 \,, t = \ln \sigma \,, \tag{5}$$

where $\sigma = \langle \overline{\chi}\chi \rangle$ and $c > 0$. Solving this equation one finds that $\lambda \sim |\ln \sigma|^{-1}$ as $\sigma \to 0$. For a Yukawa coupling $g$ one obtains in leading order [11]

$$\frac{dg}{dt} = c' g^3 \,, \quad c' > 0 \,. \tag{6}$$

Hence one expects on the basis of the triviality scenario that $g$ vanishes like

$$g \sim |\ln \sigma|^{-1/2} \,. \tag{7}$$

In order to measure effective Yukawa couplings and to compare them with (7) we have calculated the three-point function

$$G_3^{(\alpha)}(\vec{p}, \vec{q}; t_0, t_1) = \sum_{\vec{x}, \vec{x}'} e^{2i\vec{q}\cdot\vec{x}' - 2i\vec{p}\cdot\vec{x}} \\
\times \langle \overline{\chi}(2\vec{x}, t_1) M_\alpha(\vec{p} - \vec{q}, 0) \chi(2\vec{x}', t_0) \rangle_c \,. \tag{8}$$

Here $\langle \cdots \rangle_c$ denotes the fermion-line connected part of the expectation value, the lattice vectors $\vec{x}, \vec{x}'$ label spatial cubes of size $2^3$, and $M_\alpha(\vec{p}_0, t)$ is the standard pseudoscalar ($\alpha = P$) or scalar ($\alpha = S$) "meson" operator:

$$M_\alpha(\vec{p}_0, t) = \sum_{\vec{y}} e^{i\vec{p}_0 \cdot \vec{y}} \varphi_\alpha(\vec{y}) \overline{\chi}(\vec{y}, t) \chi(\vec{y}, t) \tag{9}$$

with $\varphi_P(\vec{y}) = (-1)^{y_1+y_2+y_3}, \varphi_S(\vec{y}) = 1$. In our calculations, the spatial fermion momenta $\vec{p}, \vec{q}$ take the values $(\vec{p}, \vec{q}) = (\vec{0}, \vec{0})$, $(\frac{2\pi}{L}\vec{e}_3, \vec{0})$, $(\frac{2\pi}{L}\vec{e}_3, \frac{2\pi}{L}\vec{e}_3)$ with the lattice size $L = 12$. The spatial "meson" momentum is then given by $\vec{p}_0 = \vec{p} - \vec{q}$.

In order to define effective Yukawa couplings $g_\alpha$ we compare the Monte Carlo data for the three-point functions with tree-level formulas derived from an effective lattice action describing the interaction between staggered fermions $\chi, \overline{\chi}$ and a (pseudo-)scalar field $\Phi$ via coupling terms of the form

$$-g_S \sum_x \Phi(x) \overline{\chi}(x) \chi(x) \quad \text{(scalar)} \,, \tag{10}$$

$$-g_P \sum_x \Phi(x) \overline{\chi}(x) \chi(x) (-1)^{x_1 + \cdots + x_4} \tag{11}$$
$$\text{(pseudoscalar)} \,.$$

In the tree-level formulas for the three-point functions we replace the free propagators by the full propagators calculated in the simulations, include the appropriate wave function renormalizations $Z_F$ (for the fermion field) and $Z_\alpha$ (for the "mesons") and Fourier transform also the $t_0, t_1$ dependences to obtain the following definition of the momentum-dependent Yukawa couplings $g_\alpha(p, q)$:

$$\sum_{t_1, t_0} e^{-ip_4 t_1 + iq_4 t_0} G_3^{(\alpha)}(\vec{p}, \vec{q}; t_0, t_1)$$
$$= g_\alpha(p, q) Z_F(\vec{p})^{-\frac{1}{2}} Z_F(\vec{q})^{-\frac{1}{2}} Z_\alpha(\vec{p}-\vec{q})^{-\frac{1}{2}}$$
$$\times \frac{(-8)}{V_3} \sum_t \theta_\alpha(t) e^{i(p_4 - q_4)t}$$
$$\times \frac{1}{V_3} \langle M_\alpha(\vec{q}-\vec{p}, t) M_\alpha(\vec{p}-\vec{q}, 0) \rangle_c$$
$$\times \sum_{\vec{\omega}} \overline{\theta}_\alpha(\vec{\omega}) e^{-i(\vec{p}-\vec{q})\cdot\vec{\omega}} \tag{12}$$
$$\times \Big\{ \sum_{t_0} \eta_4(\vec{\omega})^{t_0} e^{iq_4 t_0}$$
$$\times \sum_{\vec{x}, \vec{x}'} e^{2i\vec{q}\cdot(\vec{x}-\vec{x}')} \langle \chi(2\vec{x}+\vec{\omega}, t_0) \overline{\chi}(2\vec{x}', 0) \rangle \Big\}$$
$$\times \Big\{ \sum_{t_1} \eta_4(\vec{\omega})^{t_1} e^{i(p_4+\pi)t_1}$$
$$\times \sum_{\vec{x}, \vec{x}'} e^{2i\vec{p}\cdot(\vec{x}-\vec{x}')} \langle \chi(2\vec{x}+\vec{\omega}, t_1) \overline{\chi}(2\vec{x}', 0) \rangle \Big\}^* \,,$$

where $\theta_S(t) = 1$, $\theta_P(t) = (-1)^t$, $\overline{\theta}_S(\vec{\omega}) = \eta_4(\vec{\omega})$, $\overline{\theta}_P(\vec{\omega}) = 1$. The spatial volume of the lattice is denoted by $V_3$ and $\vec{\omega}$ runs over the 8 three-vectors with components 0,1. The wave function renormalizations $Z_F, Z_\alpha$ are defined in the usual way by comparing the one-particle contributions in the full propagators with the corresponding free propagators.

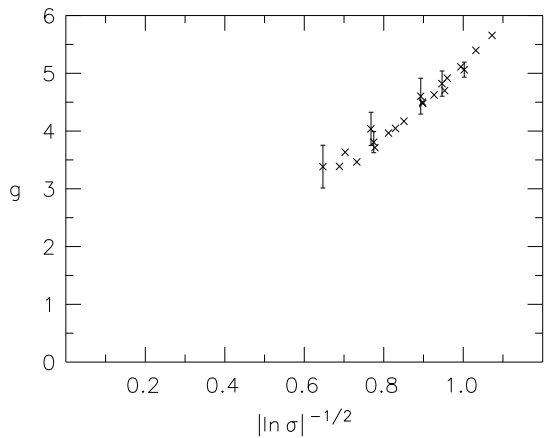

Figure 1. Effective Yukawa coupling for the Goldstone boson.

Let us now discuss the numerical results. It turns out that $g_\alpha(p,q)$ is essentially real: The imaginary part is relatively small and noisy. Hence we shall ignore it in the following and treat $g_\alpha$ as a real quantity. In order to obtain a final unique answer one would like to send the momenta to zero. However, due to the antiperiodic temporal boundary conditions for the fermions we can only give the "mesons" momentum zero by choosing $p = q$. Furthermore, we may take $\vec{p} = \vec{q} = \vec{0}$. Instead of considering the couplings at $|q_4| = \pi/L$, the smallest possible value, we eventually average $g_\alpha((\vec{0},q_4),(\vec{0},q_4))$ over $q_4$ encouraged by the observation that the data for this quantity show hardly any dependence on $q_4$. For the error we take the largest error of a single value.

This preliminary analysis leads to the data for $g_P$ plotted in fig. 1 versus $|\ln \sigma|^{-1/2}$ as suggested by eq. (7). They are consistent with a linear extrapolation to zero thereby confirming the triviality scenario. The data for $g_S$ lead to the same conclusion although less convincingly. This could be due to the fact that we have neglected the fermion-line disconnected contributions to the three-point functions, which are very difficult to calculate. Fig. 1 also demonstrates that our data are still quite far away from the continuum limit: The extrapolation has to bridge a considerable gap thus leaving room for speculations about non-trivial behaviour.

## ACKNOWLEDGEMENTS

This work was supported in part by the Deutsche Forschungsgemeinschaft. The numerical computations were performed on the Cray Y-MP in Jülich with time granted by the Scientific Council of the HLRZ. We wish to thank both institutions for their support.